\documentclass[12pt]{article}
\usepackage{graphics, latexsym}
\usepackage{graphicx}
\usepackage{setspace}
\usepackage{amssymb}
\usepackage{amsmath}
\usepackage{epsfig}
\usepackage{CJK}
\usepackage{verbatim}
\usepackage{amsmath, amssymb}
\usepackage{makecell}
\usepackage{multirow}
\usepackage{graphicx}
\usepackage{subfigure}
\usepackage{epstopdf}

\usepackage{amssymb}
\usepackage{amssymb}
\usepackage{lineno,hyperref}
\usepackage{url}
\usepackage{bm}
\usepackage{overpic}
\usepackage{graphicx}
\usepackage{mathrsfs}
\usepackage{amsfonts}
\usepackage{amssymb}
\usepackage{amsmath,amsthm}
\usepackage{indentfirst}
\usepackage{footmisc}
\usepackage{multirow}
\usepackage{xcolor}
\usepackage{tikz}
\usepackage{bm}
\usepackage{algorithm}
\usepackage{algpseudocode}
\usepackage{amsmath}
\usepackage{colortbl}
\usepackage{amsmath}

\usepackage{tabularx}
\usepackage{array}

\usetikzlibrary{arrows,shapes,chains}
\usepackage{pgf}
\usepackage{threeparttable}
\usepackage{diagbox}
\usetikzlibrary{arrows, decorations.pathmorphing, backgrounds, positioning, fit, petri, automata}
\usepackage{tikz}
\usepackage{tabu}
\usepackage{float}
\usetikzlibrary{shadows,arrows,positioning,shapes.geometric} 
\pgfdeclarelayer{background}
\pgfdeclarelayer{foreground}
\pgfsetlayers{background,main,foreground}

\usepackage{booktabs}
\usepackage{array, caption, threeparttable}
\usepackage[font=small,labelfont=bf,labelsep=none]{caption}
\newcolumntype{C}[1]{>{\centering}p{#1}}
\usepackage{makecell}

\begin{document}
\title{An CUSUM Test with Observation-Adjusted Control Limits in Change Detection }
\author{Fuquan Tang, Dong Han$^{*}$\\
Department of Statistics, Shanghai Jiao Tong University,\\
Shanghai 200030, P. R. China.\\}
\maketitle
\begin{abstract}
In this paper, we not only propose an new optimal sequential test of sum of logarithmic likelihood ratio (SLR)  but also present the CUSUM sequential test (control chart, stopping time) with the observation-adjusted control limits (CUSUM-OAL) for monitoring quickly and  adaptively the change in distribution of a sequential observations. Two limiting relationships  between the optimal test and a series of the CUSUM-OAL tests  are established. Moreover, we give the estimation of the in-control and the out-of-control average run lengths (ARLs) of the CUSUM-OAL test. The theoretical results are illustrated by numerical simulations in  detecting mean shifts of the observations sequence.
\end{abstract}

\renewcommand{\thefootnote}{\fnsymbol{footnote}}
\footnotetext{$^{*}$Supported by National Natural Science Foundation of China (11531001)
\newline$^{*}$Corresponding author, E-mail: donghan@sjtu.edu.cn
\newline $\,\, MSC\, 2010\,subject\,classification $. Primary 62L10;
Secondary 62L15}
\emph{Keywords}: Optimal sequential test, CUSUM-OAL test, change detection.

\section{INTRODUCTION}

 In order to quickly detect a change in  distribution of observations sequence without exceeding a certain false alarm rate,   a great variety of sequential tests have been proposed, developed and applied to various fields since Shewhart (1931) proposed a control chart method, see, for example, Siegmund (1985),  Basseville and Nikiforov (1993), Lai (1995, 2001), Stoumbos \emph{et al}. (2000), Chakraborti \emph{et al}. (2001),  Bersimis \emph{et al}. (2007),  Montgomery (2009), Qiu (2014), Tartakovsky \emph{et al}. (2015), Woodall \emph{et al}. (2017), Bersimis \emph{et al}. (2018) and Chakrabortia and Graham (2019).

One of popular used  sequential tests  is the following upper-sided CUSUM test which was proposed by Page (1954).
\begin{equation}
T_C(c)=\min \{n\geq 0:\,\, \max_{1\leq k\leq n}\sum_{i=n-k+1}^{n}Z_i
\geq  c\},
\end{equation}
where $c>0$ is a constant control limit, $Z_i=\log [p_{v_1}(X_i)/p_{v_0}(X_i)]$, $p_{v_0}(x)$ and $p_{v_1}(x)$ are pre-change and post-change probability density functions respectively for a sequence of mutually independent observations $\{X_i,\, i\geq 1\}$, that is, there is a unknown change-point $\tau \geq 1$ such that $X_1, ..., X_{\tau-1}$ have the probability density function $p_{v_0}$, whereas, $X_{\tau}, X_{\tau+1}, ...$ have the probability density function $p_{v_1}$. By the renewal property of the CUSUM test $T_C$ we have $sup_{k\geq 1}\textbf{E}_k(T_C-k+1|T_C\geq k)=\textbf{E}_1(T_C)$ (see Siegmund 1985, P.25), where $\textbf{E}_1(T_C)$ is the out-of-control average run length (ARL$_1$), $\textbf{P}_k$ and $\textbf{E}_k$ denote the probability and expectation respectively when the change from $p_{v_0}$ to $p_{v_1}$ occurs at the change-point $\tau=k$ for $k\geq 1$.

Though we know that the CUSUM test is optimal under Lorden's measure (see Moustakides 1986 and Ritov 1990),  the out-of-control ARL$_1$ of the CUSUM test is not small, especially in detecting small mean shifts ( see Table 1 in Section 4). In other words, the CUSUM test is insensitive in detecting  small mean shifts. Then, how to increase the sensitivity of
the CUSUM test ? Note that the control limit in the CUSUM
test is a constant $c$ which does not depend on the observation samples.
Intuitively, if the control limit of the CUSUM test can become low as the samples mean of the observation sequence increases, then the alarm time of detecting the increasing mean shifts  will be greatly shortened. Based on this idea, by selecting a decreasing function $g(x)$  we  may define the ( upper-sided ) CUSUM chart  $T_C(cg)$ with  the observation-adjusted control limits $cg(\hat{Z}_n)$ ( abbreviated to the CUSUM-OAL chart ) in the following
\begin{equation}
T_C(cg)=\min \{n\geq 0:\max_{0\leq k\leq n}\sum_{i=n-k+1}^{n}Z_i
\geq cg(\hat{Z}_n)\},
\end{equation}
where $c>0$ is a constant and $\hat{Z}_n=\sum_{i=1}^nZ_i/n$.  In other words, the control limits $cg(\hat{Z}_n)$ of the CUSUM-OAL test  can be adjusted adaptively according to the observation information $\{\hat{Z}_n\}$.  Note that the control limits $cg(\hat{Z}_n)$  may be negative. In the special case, the CUSUM-OAL chart $T_C(cg)$ becomes into the conventional CUSUM chart $T_C(c)$ in (1) when $g\equiv 1$.   Similarly, we can define a down-sided CUSUM-OAL test. In this paper,
we consider only the upper-sided CUSUM-OAL test since the properties of  the down-sided CUSUM-OAL test can be obtained by the similar method.


The main purpose of the present paper  is to show the good detection performance of the CUSUM-OAL test and to give the estimation of its the in-control and out-of-control ARLs.

The paper is organized as follows. In Section 2, we first present an optimal SLR sequential test, then define two  sequences of the CUSUM-OAL tests  and  prove that one of the two sequences of  CUSUM-OAL tests  converges to the optimal test, another sequences of  CUSUM-OAL tests converges to a combination of the optimal test and the CUSUM test.  The estimation of the in-control and out-of-control ARLs of the CUSUM-OAL tests  and their comparison are given in Section 3. The detection performances of the three CUSUM-OAL tests and the conventional CUSUM test  are illustrated in Section 4 by comparing  their numerical out-of-control ARLs. Section 5 provides some concluding remarks. Proofs of the theorems are given in the Appendix.

\section{ AN OPTIMAL SLR TEST, TWO CUSUM-OAL TESTS AND THEIR LIMITING RELATIONSHIPS}

Let $\textbf{P}_0$ and $\textbf{E}_0$ denote the probability and the expectation respectively with the probability density $p_{v_0}$  when there is no change for all the time. It is known that $\textbf{E}_{0}(Z_1)=\textbf{E}_{v_0}(\log [p_{v_1}(X_1)/p_{v_0}(X_1)]) < 0$ for $p_{v_1}(x) \neq p_{v_0}(x)$. Without loss of generality, let $\mu_0=\textbf{E}_{0}(Z_1)<0$.

It follows from Proposition 2.38 in Siegmund (1985) and (5.8)-(5.9) in Chow \emph{et al}, P.108) that the following  sequence test of sum of logarithmic likelihood ratio (SLR)
\begin{eqnarray}
T_{SLR}&=&\min\{n\geq 1:\, \prod_{k=1}^n\frac{p_{v_1}(X_k)}{p_{v_0}(X_k)}\, \geq B\,\}=\min\{n\geq 1:\, \sum_{j=1}^nZ_j \geq c\}\\
&=&\min\{n\geq 1: \, \sum_{j=1}^n(Z_j-\mu_0) \geq c+n|\mu_0|\}
\end{eqnarray}
for $B> 1$, is optimal in the following sense
\begin{eqnarray*}
\min_{T:\,\, \textbf{P}_0(T<\infty)\geq \alpha }\{\textbf{E}_1(T)\}=\textbf{E}_1(T_{SLR})
\end{eqnarray*}
for $\textbf{P}_0(T_{SLR}<\infty)=\alpha$,  where  $c=\log B$ and $0<\alpha<1$.

In particular, if $\textbf{P}_0$ is the standard normal distribution with mean shift $\mu >0$ after change-point,  we have $Z_j-\mu_0=\mu X_j$, where $\mu_0=-\mu^2/2$. It follows from proposition 4 in Fris\'{e}n (2003) that the SLR test $T_{SLR}$ in (4) is also optimal (minimal ARL$_1$) with the same false alarm probability $\textbf{P}_0(T<\tau)$.

It can be seen that the in-control average run length of $T_{SLR}$ is infinite, that is,  ARL$_0=\textbf{E}_0(T_{SLR})=\infty$. However,  the minimal ARL$_1$ with finite ARL$_0$ is a widely used optimality criterion in statistical quality control (see Montgomery, 2009) and detection of abrupt changes (see Basseville and Nikiforov, 1993). In order to get finite ARL$_0$ for $T_{SLR}$, we replace the constant control limit $c$ of $T_{SLR}$ in (3) or (4) with the dynamic control limit $n(\mu_0-r)$ and obtain a modified SLR test $T_{SLR}(r)$ in the following
\begin{equation}
T_{SLR}(r)=\min\{n\geq 1: \, \sum_{j=1}^nZ_j \geq  n(\mu_0-r)\}=\min\{n\geq 1: \, \sum_{j=1}^n(Z_j-\mu_0)\geq  -rn\}
\end{equation}
for $r\geq 0$.

For comparison, the in-control $\textbf{ARL}_{0}$ of all candidate sequential tests are constrained to be
equal to the same desired level of type I error,  the test with the lowest out-of-control $\textbf{ARL}_{v}$  has the  highest power or the fastest monitoring (detection) speed.

In the following example 1, the numerical simulations of the out-of-control ARLs of the CUSUM-OAL tests $T_C(cg_{u, 0})$ in detecting the mean shifts of observations with normal distribution will be compared with that of the SLR tests  $T^*(r)$ and $T^*(0)$, and that of the CUSUM-SLR test $T_{C}(c)\wedge T^*(0):=\min\{T_C(c), \, T^*(0)\}$ in the following Table 1. These comparisons lead us to guess that there are some limiting relationships between $T_C(cg_{u, r})$ and $T^*(r)$,  and $T_C(c\widetilde{g_{u}})$ and $T_{C}(c)\wedge T^*(0)$, respectively.

\textbf{Example 1.} Let $X_1, X_2,....$ be mutually independent following the normal
distribution $N(0, 1)$ if there is no change.  After the
change-point $\tau =1$, the mean $\textbf{E}_{\mu}(X_k)$ ( $k\geq 1$ ) will change from
$v_0=0$ to $v= 0.1, 0.25, 0.5, 0.75, 1,
1.25, 1.5, 3$.  Here, we let $p_{v_0}(x)=e^{-x^2/2}/\sqrt{2\pi}$, $p_{v_1}(x)=e^{-(x-1)^2/2}/\sqrt{2\pi}$ and therefore, $Z_k=X_k-1/2$ for $k\geq 1$, where $v_1=1$ is a  given reference value  which for the CUSUM test is the magnitude of a shift in the process mean to be detected quickly.  We conducted the numerical simulation based on 1,000,000 repetitions.

The following Table 1 lists the simulation results of the ARLs of the tests  $T_C(c)$, $T_C(c\widetilde{g_{u}})$ for $u=1, 10, 10^2, 10^3, 10^4$,  $T^*(0.0007)$,  $T_{C}(c)\wedge T^*(0)$ and $T^*(0)$  for detecting the mean shifts, where the mean shift $0.0$ means that there is no change which corresponds to  the in-control ARL$_0$ and all tests  have the common ARL$_0\approx 1000$ except the test $T^*(0)$ which has ARL$_0=\infty$. The values in the parameters are the standard deviations of the tests.

\bigskip
\bigskip
\par
\par
\footnotesize \setlength{\tabcolsep}{4pt}
\begin{tabular}{l|c|ccccccccc}
\multicolumn{11}{l}{\normalsize \textbf{Table 1}. \,\,\,\,\,\,\,\,\,\,\, \textbf{ARLs} of $T_C(c)$,  $T_C(c\widetilde{g_{u}})$, $T^*(r)$, $T_{C}(c)\wedge T^*(0)$ and $T^*(0)$ when $\tau=1$.}\\[5pt]
\hline
\textbf{Tests}& $\textbf{u, r}$      &          &       &             &\textbf{Shifts}&             &                          &             \\
\textbf{} & $\textbf{c}$&\textbf{0.0}&\textbf{0.1}&\textbf{0.25}&\textbf{0.5} &\textbf{0.75} &\textbf{1.0} &\textbf{ 1.5}&\textbf{3.00}\\
\hline
$T_C(c)$&$ $&1000.59&439.00&147.72&38.91&17.32&10.50&5.82&2.61\\
\raisebox{1.0ex}[0pt]{$ $}&c=5.0742&(993.16)&(431.89)&(140.25)&(31.79)&(11.23)&(5.49)&(2.26)&(0.66)\\
\hline
&$u=1.0$&1000.43&237.34&46.84&10.02&4.39&2.65&1.52&1.03\\
&$c=5.6125$&(1510.55)&(391.64)&(79.68)&(15.37)&(5.72)&(2.90)&(1.10)&(0.14)\\
\cline{2-11}
&$u=10.0$&999.17&18.16&4.57&2.19&1.57&1.31&1.11&1.01\\
&$c=7.7790$&(4969.50)&(77.77)&(11.06)&(3.01)&(1.45)&(0.86)&(0.38)&(0.04)\\
\cline{2-11}
&$u=10^2$&1000.66&8.15&3.38&1.92&1.47&1.27&1.11&1.01\\
&$c=9.97$&(13951.15)&(29.90)&(6.98)&(2.33)&(1.20)&(0.73)&(0.34)&(0.04)\\
\cline{2-11}
&$u=10^3$&1001.49&7.56&3.28&1.89&1.46&1.26&1.11&1.01\\
&$c=11.38$&(25423.13)&(26.45)&(6.60)&(2.27)&(1.18)&(0.72)&(0.33)&(0.04)\\
\cline{2-11}
&$u=10^4$&1001.32&7.51&3.26&1.89&1.45&1.24&1.11&1.01\\
\raisebox{10.0ex}[0pt]{$T_C(c\widetilde{g_{u}})$}&$c=11.84$&(30042.90)&(26.22)&(6.56)&(2.27)&(1.17)&(0.72)&(0.33)&(0.04)\\
\hline
&$r=0.0007$&1001&7.44&3.25&1.88&1.45&1.25&1.09&1.01\\
\raisebox{1.0ex}[0pt]{$T^*(r)$}& &(43017)&(26.07)&(6.57)&(2.24)&(1.18)&(0.71)&(0.33)&(0.04)\\
\hline
\cline{2-11}
&$r=0$&1001.13&7.52&3.25&1.88& 1.45&1.25&1.09&1.01\\
\raisebox{1.0ex}[0pt]{$T_{C}(c)\wedge T^*(0)$}&$c=11.9271$&(31090.71)&(26.88)&( 6.57)&(2.24)&(1.17)&(0.72)&(0.33)&(0.04)\\
\hline
\cline{2-11}
&$r=0$& $\infty$ &7.47&3.25&1.88&1.45&1.25&1.09&1.01\\
\raisebox{1.0ex}[0pt]{$T^*(0)$}& &($\infty$)&(26.24)&(6.54)&(2.27)&(1.18)&(0.72)&(0.33)&(0.04)\\
\hline \multicolumn{11}{l}{}
\end{tabular}
\normalsize

From the last row in Table 1, it's a little surprising that though  the ARL$_0$ of $T^*(0)$ is infinite, that is, $\textbf{E}_0(T^*(0))=\infty$, the detection speed of  $T^*(0)$ is  faster than that of the CUSUM chart $T_C$ for all mean shifts, in particular, for detecting the small mean shift 0.1, the speed of $T^*(0)$ is only 7.47 which is  very faster than the speed, 439, of the CUSUM test. Moreover, both control charts $T^*(0.0007)$ and $T_{C}(11.9271)\wedge T^*(0)$ not only have the nearly same detection performance as $T^*(0)$ but also can have the finite in-control ARL$_0$. Note particularly that when the number $u$ in $\widetilde{g_{u}}$ is taken from $0$ to $1, 10, 10^2, 10^3, 10^4$, the detection speed of $T_C(c\widetilde{g_{u}})$ is getting faster and faster,  approaching to that of $T_{C}(c)\wedge T^*(0)$. This inspires us to prove the following theoretic results.

\textbf{Theorem 2.}  \textit{ Let  $\tau=1$ and $\{X_k, k\geq 1\}$ be an i.i.d. observations sequence with $\mu_0=\textbf{E}_{0}(Z_1)<0$. Then
\begin{eqnarray}
\lim_{u\to \infty}T_C(cg_{u, r})=T^*(r),\,\,\,\,\,\,\,\, \lim_{u\to
\infty}T_C(c\widetilde{g_{u}})=T_{C}(c)\wedge T^*(0).
\end{eqnarray}}

Theorem 2 shows that when the constant control limit $c$  of the CUSUM test $T_C(c)$  is replaced with the observation-adjusted control limits $\{cg_{u, r}(\hat{Z}_n)\}$ and $\{c\widetilde{g_{u}}(\hat{Z}_n)\}$ respectively, the corresponding two CUSUM-OAL tests $\{T_C(cg_{u, r})\}$ and $\{ T_C(c\widetilde{g_{u}})\}$ will converge to the optimal SLR test  $T^*(r)$ and the CUSUM-SLR test $T_{C}(c)\wedge T^*(0)$ as $u\to \infty$, respectively. In other words, the fastest alarm times that $\{T_C(cg_{u, r})\}$ and $\{ T_C(c\widetilde{g_{u}})\}$ can be reached are $T^*(r)$ and $T_{C}(c)\wedge T^*(0)$, respectively.

\textbf{Remark 2.} Since $T_C(cg_{0, r})=T_C(c\widetilde{g_{0}})=T_C(c)$ when $u=0$, it follows that both $\{T_C(cg_{u, r}):\, u\geq 0\}$ and $\{ T_C(c\widetilde{g_{u}}):\, u\geq 0\}$ can be seen as two "long bridges" connecting $T_C(c)$ and $T^*(r)$, and $T_C(c)$ and $T_{C}(c)\wedge T^*(0)$, respectively.

\section{ESTIMATION  AND COMPARISON OF ARL OF THE CUSUM-OAL TEST}

In this section we will give  an estimation  of the  ARLs of the following  CUSUM-OAL test that can be written as
\begin{equation}
T_C(cg)=\min \{n\geq 1:\max_{1\leq k\leq n}\sum_{i=n-k+1}^{n}Z_i
\geq cg(\hat{Z}_n(ac))\}, \label{f07}
\end{equation}
where $g(.)$ is a decreasing function,
$\hat{Z}_n(ac)=\frac{1}{j}\sum_{i=n-j+1}^nZ_i, j=\min\{n, [ac]\}$
for $ac \geq 1, a>0, c>0$,  and $[x]$ denotes the smallest integer
greater than  or equal to $x$. Here $\hat{Z}_n(ac)$ is a sliding
average of the statistics, $Z_i, n-j+1\leq i\leq n$,  which will become
$\hat{Z}_n=\frac{1}{n}\sum_{i=1}^nZ_i$ when $a=\infty$.

Next we discuss on the the post-change probability distribution in
order to estimate the ARLs of  $T_C(cg)$.

Usually we rarely know the post-change probability distribution
$P_{v}$ of the observation process before it is detected. But the
possible change domain   and its boundary (including the size and
form of the boundary) about $v$  may be determined by engineering
knowledge, practical experience or statistical data. So we may
assume that the region of parameter space $V$ and a probability
distribution $Q$ on $V$ are known.  If we have no prior knowledge of
the possible value of $v$ after the change time $\tau$,  we may
assume that $v$ occurs equally on $V$,  that is, the probability
distribution $Q$ is an  equal probability distribution (or uniform
distribution ) on $V$.  For example, let $P_{v}$ be the normal
distribution and $v=(\mu, \sigma)$, where $\mu$ and $\sigma $ denote
the mean and standard deviation respectively, we can take the set
$V=\{(\mu, \sigma): \mu_1 \leq \mu \leq \mu_2, 0< \sigma_1\leq
\sigma \leq \sigma_2\}$ and $Q$ is subject to the uniform
distribution $U(V)$ on $V$ if $v$ occurs equally on $V$, where the
numbers $\mu_1, \mu_2, \sigma_1$ and $\sigma_2$ are known. It means
that we know the domain of the possible post-change distributions,
$P_v, v\in V$, i.e., the boundary $\partial{V}$ of the parameter
space $V$ is known.


Next we shall divide the parameter space $V$ into three subsets
$V^{+}$, $V^0$ and $V^{-}$ by the  Kullback-Leibler information
distance.  Let
\begin{eqnarray*}
V^{-}=\{v: E_{v}(Z_1)<0\},\,\,\,\,\,\,V^{0}=\{v:
E_{v}(Z_1)=0\},\,\,\,\,\,\,V^{+}=\{v: E_{v}(Z_1)>0\}
\end{eqnarray*}
where $E_{v}(Z_1)=I(P_{v}|P_{v_0})-I(P_{v}|P_{v_1})$ and
\begin{eqnarray*}
I(P_{v}|P_{v_0})=E_{v}(\log[\frac{P_{v}(X_1)}{P_{v_0}(X_1)}]),\,\,\,\,\,\,\,\,
I(P_{v}|P_{v_1})=E_{v}(\log[\frac{P_{v}(X_1)}{P_{v_1}(X_1)}])
\end{eqnarray*}
are two Kullblak-Leibler information distances  between $P_{v}$,
$P_{v_0}$ and $P_{v}$, $P_{v_1}$. Since $I(p|q)= 0$ if and only if
$p=q$, where $p$ and $q$ are two probability measures, it follows
that $E_{v_0}(Z_1)=-I(P_{v_0}|P_{v_1})$, and therefore, $v_0 \in
V^{-}$ when $P_{v_1}\neq P_{v_0}$. When $v \in V^{-}$, i.e.,
$I(P_{v}|P_{v_0})<I(P_{v}|P_{v_1})$, it means that $P_{v}$ is closer
to $P_{v_0}$ than to $P_{v_1}$ according to the Kullblak-Leibler
information distance. There is a similar explanation for $v \in
V^{+}$ or $\in V^{0}$.

Suppose the post-change distribution $P_{v}$ and the function
$g(x)$ satisfy the following conditions:\\
\textbf{(I)} The probability $P_{v}$ is not a point mass at $E_{v}(Z_1)$ and $P_v(Z_1>0)>0$.\\
\textbf{(II)} The  moment-generating function
$h_{v}(\theta)=E_{v}(e^{\theta Z_1})$ satisfies
$h_{v}(\theta)<\infty $ for some $\theta >0$. \\
\textbf{(III)} The function  $g(x)$ is decreasing, its second order
derivative function $g''(x)$ is continuous and bounded, and there is a positive number $x^*$ such that $g(x^*)=0$.

Let $\tilde{Z}_1=Z_1+|g'(\mu)|(Z_1-\mu)/a$,
$\tilde{h}(\theta)=\textbf{E}_v(e^{\theta \tilde{Z}_1})$ and
\begin{eqnarray*}
H_v(\theta)=\frac{au}{g(\mu)}\ln \tilde{h}_v(\theta) +
(1-\frac{au}{g(\mu)})\ln h_{v}(\theta)
\end{eqnarray*}
for $\theta \geq 0$, where $E_{v}(Z_1)=\mu <0$, $a\leq g(\mu)/u$ and
$u=H'_v(\theta^*_v)$, where $\theta^*_v >0$ satisfies
$H_v(\theta^*_v)=0$.  Note that $H_v(\theta)$ is a convex function
and $H'_v(0)=\mu<0$. It follows that there is a unique positive
number $\theta^*_v >0$ such that $H_v(\theta^*_v)=0$.


Note that the following function
\begin{eqnarray*}
\Theta(x)=\theta(\frac{1}{x})-xH_{v}(\theta(\frac{1}{x}))-2\theta^*_v
\end{eqnarray*}
satisfies that $\Theta(1/u)=\theta(u)-2\theta^*_v=-\theta^*_v$,
where $\theta(u)=\theta^*_v$,
\begin{eqnarray*}
\Theta'(x)=-\frac{1}{x^2}\theta'(\frac{1}{x})-H_{v}(\theta(\frac{1}{x}))+\frac{1}{x}\theta'(\frac{1}{x})H_v'(\theta(\frac{1}{x}))=-H_{v}(\theta(\frac{1}{x}))
\end{eqnarray*}
and therefore,  $\Theta'(\theta(u))=-H(\theta(u))=-H(\theta^*_v)=0$,
$\Theta'(\theta(1/x))>0$ for $x>1/u$ and $\Theta'(\theta(1/x))<0$
for $x>1/u$. Hence, there exists a positive number $b$ defined in
(\ref{f08}).

It can be seen, the main part of $\textbf{ARL}_{v}(T_{c}(g))$ will
be an exponential function, square function, and linear function of
$c$ when  the process $\{Z_k: k\geq 0\}$ has  no change or a "small
change", a "medium change"  and  a "large change" from
$P_{v_0}$ to $P_{v}$, respectively.  Here, the "small change"
($v\in V^-$) means that $P_{v}$ is closer to $P_{v_0}$ than to
$P_{v_1}$, i.e., $I(P_{v}|P_{v_0})<I(P_{v}|P_{v_1})$, and the
"large change" is just the opposite.  The "medium change" ($v\in
V^0$) corresponds to $I(P_{v}|P_{v_0})=I(P_{v}|P_{v_1})$.

%
%
%

In this paper, we will use another  method to prove Theorem 3
since Wald's identity and the martingale method do not hold or can
not work for showing the  ARLs estimation of the test $T_c(g)$ when
$g$ is not constant.

Next we compare the detection performance of the CUSUM-OAL  test
($\textbf{ARL}_{v}(T_{c'}(g))$)  with that of  the CUSUM test
($\textbf{ARL}_{v}(T_{C}(c))$) by using (\ref{f11}) in Theorem 4.1.

Let $\textbf{ARL}_{v_0}(T_{c'}(g))=\textbf{ARL}_{v_0}(T_{C}(c))$ for
large $c'$ and $c$. We have $c=c'\theta_{v_0}^*g(\mu_0) +o(1)$.
Hence
\begin{eqnarray*}
\textbf{ARL}_{v}(T_{C}(c))>\textbf{ARL}_{v}(T_{c}(g))
\end{eqnarray*}
for $s^*_{v}\theta^*_{v_0}> g(\mu)\theta^*_v/g(\mu_0)$ when
$\mu_0<\mu <0$ and for $\theta^*_{v_0}>g(\mu)/g(\mu_0)$ when
$\mu\geq 0$.  This means that $\textbf{ARL}_{v}(T_{c}(g))$ can be
smaller than $\textbf{ARL}_{v}(T_{C}(c))$ as long as
$g(\mu)/g(\mu_0)$ is small for all $\mu >\mu_0$.

\bigskip

\section{ NUMERICAL SIMULATION AND A REAL EXAMPLE ILLUSTRATION }

\subsection{Numerical Simulation of ARLs for $\tau$$ \geq 1$}

By the simulation results of ARLs in Table 1, we see that the detection performance of  $T^*(r)$, $T_{C}(c)\wedge T^*(0)$, $T^*(0)$ and $T_C(c\widetilde{g_{u}})$ for large $u$ is  much better than that of the conventional CUSUM test $T_C$ for  $\tau=1$.

The following Table 2 illustrates the simulation values of $\textbf{E}_{\tau_i, v}$ and $\textbf{J}_{ACE} $ of nine tests in detecting two mean shifts $v=0.1$ and $v=1$ after six change-points, $\tau_i, \, 1\leq i\leq 6$ with ARL$_0(T)=\textbf{E}_0(T)\approx 500$.

\bigskip
\bigskip
\par
\par
\footnotesize \setlength{\tabcolsep}{3pt}
\begin{tabular}{l|c|ccccccccc}
\multicolumn{11}{l}{\normalsize \textbf{Table 2}. \,Simulation of $\textbf{E}_{\tau_i, v}$ and $\textbf{J}_{ACE} $ for detecting two mean shifts $v=0.1, \,v=1$.}\\[5pt]
\hline
\textbf{Tests}&\textbf{v} &$ \underset{\tau_0=0}{\textbf{E}_{0,v}} $&$ \underset{\tau_1=1}{\textbf{E}_{1,v}} $&$ \underset{\tau_2=10}{\textbf{E}_{10,v}} $& $ \underset{\tau_3=50}{\textbf{E}_{50,v}} $&$ \underset{\tau_4=100}{\textbf{E}_{100,v}} $&$ \underset{\tau_5=150}{\textbf{E}_{150,v}} $&$ \underset{\tau_6=200}{\textbf{E}_{200,v}} $&$\textbf{J}_{ACE}$ \\
\hline   
                                       &     &\emph{parameter}  &\emph{c=4.3867} &&&&&&&\\
                                       &v=0.1&498.55      &247.25     &241.73     &243.52      &243.87     &244.23      &243.28      &243.98&\\ 
$T_C(c)$                               &     &(493.25)    &(241.76)   &(243.46)   &(263.86)    &(288.53)   &(314.22)    &(339.73)    &&\\               
                                       &v=1   &498.55      &9.15       &8.48       &8.46        &8.46       &8.46        &8.46        &8.58  &\\ 
                                       &     &(493.25)    &(5.00)     &(5.13)     &(5.87)      &(6.77)     &(7.63)      &(8.49)      &&\\ 
\hline
                                       &     &\emph{parameter}  &\emph{c=6.5839} &&&&&&&\\
                                       &v=0.1&498.11      &8.06       &41.52      &95.15       &136.72     &167.20      &193.34      &107.00&\\
$T_C(c\widetilde{g}_{100})$            &     &(6540.43)   &(29.29)    &(184.62)   &(501.30)    &(781.77)   &(1013.44)   &(1225.66)   &&\\ 
                                       &v=1  &498.11      &1.25       &4.15       &9.03        &12.38      &14.69       &16.60       &9.68  &\\
                                       &     &(6540.43)   &(0.73)     &(10.64)    &(34.72)     &(56.19)    &(72.86)     &(88.02)     &&\\
\hline
                                       &     &\emph{parameter}  &\emph{r=0.00137} &&&&&&&\\
                                       &v=0.1&499.43      &7.44       &38.03      &87.34       &123.01     &150.69      &172.79      &96.55 &\\
$T^*(r)$                               &     &(16588.89)  &(26.07)    &(168.31)   &(473.17)    &(734.40)   &(966.69)    &(1164.92)   &&\\
                                       &v=1  &499.43      &1.25       &4.11       &9.02        &12.73      &15.51       &17.86       &10.08 &\\
                                       &     &(16588.89)  &(0.71)     &(10.72)    &(36.12)     &(61.00)    &(82.62)     &(102.32)    &&\\   
\hline
                                       &     &\emph{parameter}  &\emph{c=10.4889} &&&&&&&\\
$T_{C}(c)$                             &v=0.1&499.58      &7.50       &38.30      &88.61       &125.16     &152.58      &179.39      &98.59 &\\
$\wedge$                               &     &(10932.09)  &(26.43)    &(170.16)   &(476.22)    &(746.80)   &(966.95)    &(1202.64)   &&\\
$T^*(0)$                               &v=1  &499.58      &1.25       &4.11       &9.04        &12.10      &13.88       &15.10       &9.25  &\\    
                                       &     &(10932.09)  &(0.72)     &(10.70)    &(35.90)     &(56.80)    &(71.61)     &(83.38)     &&\\                                   
\hline
                                       &v=0.1&$\infty$    &7.50       &38.58      &88.64       &125.00     &152.92      &177.42      &98.34 &\\
                                       &     &($\infty$)  &(26.54)    &(170.58)   &(475.69)    &(743.32)   &(972.55)    &(1191.50)   &&\\     
\raisebox{1.0ex}[0pt]{ $T^*(0)$}       &v=1  &$\infty$    &1.25       &4.12       &9.09        &12.77      &15.60       &18.02       &10.14 &\\
                                       &     &($\infty$)  &(0.72)     &(10.75)    &(36.28)     &(60.67)    &(82.23)     &(101.86)    &&\\                                      
\hline              
                                       &         &\emph{parameter}  &\emph{m=50} &\emph{c=5.8093} &\emph{p=1} &&&&&\\
                                       &v=0.1    &500.87            &164.18      &251.68          &266.00     &266.51   &266.07   &266.44    &246.81  &\\
                                       &         &(686.03)          &(239.82)    &(348.33)        &(377.68)   &(399.06) &(418.45) &(438.60)  &&\\       
                                       &v=1      &500.87            &2.68        &7.34            &9.41       &9.34     &9.35     &9.36      &7.91    &\\
                                       &         &(686.03)          &(2.94)      &(7.25)          &(9.45)     &(10.04)  &(10.70)  &(11.36)   &&\\                                              
\cline{2-11}
                                       &         &\emph{parameter}  &\emph{m=30}      &\emph{c=6.7701}&\emph{p=1}       &&&&&\\
                                       &v=0.1    &500.64     &174.61    &259.52   &264.36   &264.87   &264.67   &264.55   &248.76  &\\
                                       &         &(666.64)   &(244.09)  &(345.62) &(366.62) &(388.06) &(407.65) &(428.29) &&\\      
                                       &v=1      &500.64     &2.80      &8.07     &9.36     &9.37     &9.36     &9.36     &8.05    &\\
                                       &         &(666.64)   &(3.18)    &(7.68)   &(8.82)   &(9.52)   &(10.18)  &(10.84)  &&\\                                       
\cline{2-11}
                                       &         &\emph{parameter}  &\emph{m=10}      &\emph{c=26.3031}&\emph{p=1}      &&&&&\\
\raisebox{10.0ex}[0pt]{$T_C(ch_m)$}    &v=0.1    &500.12     &218.56    &288.10   &287.93   &287.44   &287.59   &287.68    &276.22  &\\
                                       &         &(613.17)   &(276.99)  &(356.17) &(375.19) &(398.04) &(422.13) &(447.11)  &&\\
                                       &v=1      &500.12     &3.77      &9.35     &9.31     &9.30     &9.30     &9.30      &8.39    &\\
                                       &         &(613.17)   &(4.83)    &(7.44)   &(8.02)   &(8.73)   &(9.47)   &(10.23)   &&\\                                                                         
\hline
                                       &v=0.1    &500.38      &6.89     &32.94    &58.76    &46.86    &7.57     &88.88    &40.32 &\\
                                       &         &(15373.79)  &(19.02)  &(116.99) &(248.01) &(223.84) &(130.62) &(2208.53)&&\\                        
\raisebox{1.0ex}[0pt]{$T^*_{M}$}       &v=1      &500.38      &1.25     &4.11     &9.07     &12.79    &1.25     &9.12     &6.27  &\\
                                       &         &(15373.79)  &(0.72)   &(10.72)  &(36.27)  &(60.88)  &(6.60)   &(175.62) &&\\
\hline \multicolumn{11}{l}{}
\end{tabular}
\begin{tablenotes}
     \item Tablenotes:\emph{\textbf{ The parameters for $T^*_{M}$ are k1=1, k2=150, $r_1=5.2*10^{-5}$, $r_2=1.1*10^{-5}$, and the expectation and standard deviation in both cases are 1717.06 with 13459.80 and 3918.33 with 16893.25, respectively.}}
\end{tablenotes} 
\normalsize

\subsection{A Real Example }

\section{CONCLUSION }

The contributions of this paper can be summarized to the following
three aspects.

(1) We present the optimal test $T^*(r)$ under generalized ARL$_0$ with finite
$\textbf{E}_0(T^*(r))$.

(2) To enhance the sensitivity of the CUSUM test for detecting the
distribution change we propose a CUSUM-type test with a real-time
observation control limit (CUSUM-OAL). Numerical simulations show
that the out-control ARLs of the CUSUM-OAL tests are significantly smaller
than the out-control ARLs of the  CUSUM test. But the CUSUM-OAL tests have
bigger standard deviations than the CUSUM test in the in-control
state. Moreover, we obtain the estimations of the ARLs of the
CUSUM-OAL tests. Both theoretical estimations and numerical
simulations show that the CUSUM-OAL tests perform much better than
the CUSUM test when $\tau =1$.

%
%

\newpage
\vskip 0.2cm
\textbf{Appendix : Proofs of Theorems}
\setcounter{equation}{0}
\renewcommand\theequation{A. \arabic{equation}}
\normalsize

\bigskip

\textbf{Proof of Theorem 1.}  Let $v\in V^{-}$. We first prove that
\begin{equation}
e^{cg(\mu)\theta^*_v(1+o(1))}/bc\leq E_{v}(T_c(g))\leq
cu^{-1}g(\mu)e^{cg(\mu)\theta^*_v(1+o(1))}
\end{equation}
for a large $c$.

Next we first prove the upward inequality of (36). Let $m_1=\left\lceil
cu^{-1}g(\mu) \right\rceil$, $m_k=\left\lceil km_1\right\rceil $ for
$k\geq 0$ and  $m= \left\lceil
tm_1\exp\{cg(\mu)\theta^*_v(1+o(1))\}\right\rceil $ for $t>0$,
where $ \left\lceil x\right\rceil $ denotes the smallest integer
greater than or equal to $x$.   Without loss of generality, the
number $ \left\lceil x\right\rceil $ will be replaced by $x$ in the
following when $x$ is large.   It follows that
\begin{eqnarray}
P_{v}(T_c(g)>m)&=&P_{v}(\sum_{i=n-k+1}^{n}Z_{i}<cg(\hat{Z}_n(ac)), \
\ \ 1\leq k\leq n, 1 \leq
n\leq m) \nonumber\\
&\leq & P_{v}(\sum_{i=m_{j}-m_1+1}^{m_j}Z_{i}<cg(\hat{Z}_{m_j}(ac)), \,\, 1\leq j\leq m/m_1 )\nonumber\\
&= & [P_v(\sum_{i=1}^{m_1}Z_{i}<cg(\hat{Z}_{m_1}(ac))]^{m/m_1}
\end{eqnarray}
for a large $c$,  where
$\hat{Z}_{m_j}(ac))=(ac)^{-1}\sum_{i=m_j-ac+1}^{m_j}Z_i$ and the
last quality holds since the events
\begin{eqnarray*}
\{\sum_{i=m_{j}-m_1+1}^{m_j}Z_{i}<cg(\hat{Z}_{m_j}(ac))\},
\end{eqnarray*}
$1\leq j\leq m/m_1$, are mutually independent and have an identity
distribution.  Since $\hat{Z}_{m_1}(ac)-\mu \to 0 (a.s.)$ and
$ac(\hat{Z}_{m_1}(ac)-\mu )^2 \Rightarrow \chi^2$ ($\chi^2$
-distribution)  as $c\to \infty$, it follows that
\begin{eqnarray*}
g(\hat{Z}_{m_1}(ac))=g(\mu)+g'(\mu)(\hat{Z}_{m_1}(ac)-\mu)+O(1/c)
\end{eqnarray*}
and
\begin{eqnarray*}
P_v(\sum_{i=1}^{m_1}Z_{i}<cg(\hat{Z}_{m_1}(ac))=P_v(\sum_{i=m_1-ac+1}^{m_1}\tilde{Z}_{i}+\sum_{i=1}^{m_1-ac}Z_{i}<c(g(\mu)+O(1/c)))
\end{eqnarray*}
for a large $c$, where $\tilde{Z}_{i}=Z_i+a^{-1}|g'(\mu)|(Z_i-\mu)$.
Let
\begin{eqnarray*}
\tilde{h}_v(\theta)=E_v(e^{\theta
\tilde{Z}_1}),\,\,\,h_{v}(\theta)=E_{v}(e^{\theta Z_1}).
\end{eqnarray*}
and
\begin{eqnarray*}
H_v(\theta)=\frac{au}{g(\mu)}\ln \tilde{h}_v(\theta) +
(1-\frac{au}{g(\mu)})\ln h_{v}(\theta).
\end{eqnarray*}
Note that $H_v(\theta)$ is a convex function and $H'_v(0)=\mu<0$.
This means that there is a unique positive number $\theta^*_v >0$
such that $H_v(\theta^*_v)=0$. Let $u=H'_v(\theta^*_v).$ It follows
from (A.9) that
\begin{eqnarray*}
&&P_v(\sum_{i=m_1-ac+1}^{m_1}\tilde{Z}_{i}+\sum_{i=1}^{m_1-ac}Z_{i}\geq c(g(\mu)+O(1/c)))\\
&&=P_v(\sum_{i=m_1-ac+1}^{m_1}\tilde{Z}_{i}+\sum_{i=1}^{m_1-ac}Z_{i}\geq m_1u(1+O(1/c)))\\
&&\geq \exp\{-m_1(\theta u' -H_v(\theta)+\frac{1}{m_1}\log (F^{m_1}_{\theta}(m_1u')-F^{m_1}_{\theta}(m_1u))+o(1))\}\\
&&=\exp\{-cg(\mu)(\theta\frac{u'}{u} -\frac{1}{u}H_v(\theta)+o(1))\}
\end{eqnarray*}
for a large $c$. Taking $\theta \searrow \theta^*_v$ and $u'\searrow
u$, we have
\begin{eqnarray*}
P_v(\sum_{i=m_1-ac+1}^{m_1}\tilde{Z}_{i}+\sum_{i=1}^{m_1-ac}Z_{i}<
c(g(\mu)+O(1/c)))\leq 1-\exp\{-cg(\mu)\theta^*_v(1+o(1))\}
\end{eqnarray*}
for a large $c$. Thus, by (A.11) we have
\begin{eqnarray}
P_{v}(T_c(g)>m)\leq [ 1-\exp\{-cg(\mu)\theta^*_v(1+o(1))\}]^{m/m_1}
\to e^{-t}.
\end{eqnarray}
as $c\to \infty$. By the properties of exponential distribution, we
have
\begin{eqnarray*}
E_v(T_c(g)))\leq cu^{-1}g(\mu)e^{cg(\mu)\theta^*_v(1+o(1))}
\end{eqnarray*}
for a large $c$.

To prove the downward inequality of (A.10), let
\begin{eqnarray*}
U_m &=&\{\sum_{i=n-k+1}^{n} Z_{i}<cg(\hat{Z}_n(ac)),\;\; 1\leq k\leq  ac-1, \, bc\leq n \leq m \}\\
V_m &=&\{\sum_{i=n-k+1}^{n} Z_{i}<cg(\hat{Z}_n(ac)),\;\; ac\leq k\leq  bc-1, \, bc\leq n \leq m \}\\
W_m &=&\{\sum_{i=n-k+1}^{n} Z_{i}<cg(\hat{Z}_n(ac)), \;\; bc\leq k\leq n, \; bc\leq n\leq m \}\\
S_{bc}&=&\{\sum_{i=n-k+1}^{n} Z_{i}<cg(\hat{Z}_n(ac)), \;\; 1\leq
k\leq n, \; 1\leq n\leq bc-1 \},
\end{eqnarray*}
where $b$ is defined in (\ref{f08}) and without loss of generality,
we assume that $b>a$. Obviously, $\{T_c(g)>m\}=U_m V_m W_{m}S_{bc}$.

Let $k=xcg(\mu)$. By Chebyshev's inequality, we have
\begin{eqnarray*}
P_v(\sum_{i=n-k+1}^{n} Z_{i}<cg(\hat{Z}_n(ac)))&=&P_v\Big(\sum_{i=n-k+1}^{n} \tilde{Z}_{i}+\sum_{i=n-ac+1}^{n-k}\tilde{\tilde{Z}}_i< cg(\mu)(1+o(1))\Big)\\
&\geq &1-\exp\{-cg(\mu)(\theta -x\tilde{H}_v(\theta)+o(1))\}
\end{eqnarray*}
for $1\leq k\leq  ac-1, \, bc\leq n \leq m$, where
$\tilde{\tilde{Z}}_i=-g'(\mu)(Z_{i}-\mu)/a$ and
\begin{eqnarray*}
\tilde{H}_v(\theta)=\ln \tilde{h}_v(\theta) + (\frac{ac}{k}-1)\ln
\hat{h}_{v}(\theta),\,\,\,\,\, \hat{h}_{v}(\theta)=E_v(e^{\theta
\tilde{\tilde{Z}}_i}).
\end{eqnarray*}
Since $\tilde{H}_v(\theta)$ and  $H_v(\theta)$ are two convex
functions and
\begin{eqnarray*}
&&\tilde{H}'_v(0)-H'_v(0)=0,\\
&&\tilde{H}''_v(0)-H''_v(0)=\sigma^2[(1+\frac{g'(\mu)}{a})^2+(\frac{ac}{k}-1)-\frac{au}{g(\mu)}(1-\frac{g'(\mu)}{a})^2+\frac{au}{g(\mu)}-1]>0,
\end{eqnarray*}
it follows that $\tilde{\theta}_v^* \geq \theta_v^*$, where
$\tilde{\theta}_v^*$ and $\theta_v^*$ satisfy
$\tilde{H}_v(\tilde{\theta}_v^*)=H_v(\theta_v^*)=0$. Hence
\begin{align}
P_v(\sum_{i=n-k+1}^{n}
\tilde{Z}_{i}+\sum_{i=n-ac+1}^{n-k}\tilde{\tilde{Z}}_i&<cg(\mu)(1+o(1)))\nonumber\\
&\geq 1-\exp\{-cg(\mu)\theta_v^*(1+o(1))\}
\end{align}
for $1\leq k\leq  ac-1, \, bc\leq n \leq m$. Similarly, we can get
\begin{align}
P_v(\sum_{i=n-ac+1}^{n} \tilde{Z}_{i}+\sum_{i=n-k+1}^{n-ac}Z_{i}&<
cg(\mu)(1+o(1)))\nonumber\\
&\geq 1-\exp\{-cg(\mu)\theta_v^*(1+o(1))\}
\end{align}
for $ac\leq k\leq  bc-1, \, bc\leq n \leq m$, and
\begin{eqnarray}
P_v(\sum_{i=n-ac+1}^{n} \tilde{Z}_{i}+\sum_{i=n-k+1}^{n-ac}Z_{i}&<&
cg(\mu)(1+o(1)))\nonumber\\
&\geq& 1-\exp\{-2cg(\mu)\theta_v^*(1+o(1))\}
\end{eqnarray}
for $bc\leq k\leq  n, \, bc\leq n \leq m$.

Let $m=tcg(\mu)\theta_v^*/bc$ for $t>0$. By (A.13), (A.14), (A.15)
and Theorem 5.1 in Esary, Proschan and Walkup (1967) we have
\begin{eqnarray*}
P_v(U_mV_m)&\geq & \prod_{n=bc}^{m}\prod_{k=1}^{ bc-1}P_v(\sum_{i=n-k+1}^{n} \tilde{Z}_{i}-\frac{g'(\mu)}{a}\sum_{i=n-ac+1}^{n-k}(Z_{i}-\mu)< cg(\mu)(1+o(1)))\\
&\geq& [1-\exp\{-cg(\mu)\theta_v^*(1+o(1))\}]^{bc m} \to e^{-t}
\end{eqnarray*}
and
\begin{eqnarray*}
P_v(W_m)&\geq & \prod_{n=bc}^{m}\prod_{k=bc}^{ n}P_v(\sum_{i=n-ac+1}^{n} \tilde{Z}_{i}+\sum_{i=n-k+1}^{n-ac}Z_{i}<cg(\mu)(1+o(1)))\\
&\geq& [1-\exp\{-2cg(\mu)\theta_v^*(1+o(1))\}]^{(m-bc)^2} \to 1
\end{eqnarray*}
as $c\to +\infty$.

Finally,
\begin{eqnarray*}
P_v(S_{bc})&\geq & P_v(\sum_{i=n-k+1}^{n} Z_{i}<cg_0, \;\; 1\leq k\leq n, \; 1\leq n\leq bc-1)\\
 &\geq &\prod_{n=1}^{bc-1}\prod_{k=1}^{ n}(1-\exp\{-cg_0\theta +k\ln h_v(\theta)\})\\
&\geq & [1-\exp\{-cg_0\theta_0\}]^{(bc)^2} \to 1
\end{eqnarray*}
as $c\to +\infty$, where $\theta_0>0$ satisfies $h_v(\theta_0)=1$.

Thus
\begin{eqnarray*}
P_{v}(T_c(g)>m)=P_v(U_m V_m W_{m}S_{bc})\searrow  e^{-t}.
\end{eqnarray*}
as $c\to \infty$. This implies that
\begin{eqnarray*}
 E_{v}(T_c(g)) \geq e^{cg(\mu)\theta^*_v(1+o(1))}/bc
\end{eqnarray*}
for a large $c$. This completes the proof of (A.10).

Let $v\in V^0$.

Let $m_1=(cg(0))^2/\sigma^2$. It follows that
\begin{eqnarray*}
E_v(T_c(g))&=&\sum_{n=0}^{\infty}P_v(T_c(g)> n)\\
&\leq&  m_1+\sum_{n=m_1}^{2m_1}P_v(T_c(g)> n)+...+\sum_{n=km_1}^{(k+1)m_1}P_v(T_c(g)> n)+....\\
&\leq & m_1[1+\sum_{k=1}^{\infty}P_v(T_c(g)> km_1)].
\end{eqnarray*}
Note that
\begin{eqnarray*}
&&P_v(T_c(g)> km_1)\leq  P_v\Big(\sum_{i=(j-1)m_1+1}^{jm_1}Z_{i}+\sum_{i=jm_1-ac+1}^{jm_1}Z'_{i}<cg(0)(1+o(1)), \,\, 1\leq j \leq k \Big)\\
&=& [P_v\Big(\sum_{i=1}^{m_1}Z_{i}+\sum_{i=m_1-ac+1}^{m_1}Z'_{i}<cg(0)(1+o(1))\Big)]^k\\
&=&[P_v\Big(\sum_{i=1}^{m_1-ac}Z_{i}+\sum_{i=m_1-ac+1}^{m_1}(1+A)Z_{i}\,< cg(0)(1+o(1))\Big)]^k\\
&=&[P_v\Big(\frac{\sum_{i=1}^{m_1-ac}Z_{i}}{cg(0)}+\frac{\sum_{i=m_1-ac+1}^{m_1}(1+A)Z_{i}}{cg(0)}\,<
(1+o(1))\Big)]^k
\end{eqnarray*}
for a large $c$, where $A=|g'(0)|/a$,  and
\begin{eqnarray*}
\frac{\sum_{i=1}^{m_1-ac}Z_{i}}{cg(0)} \Rightarrow X\sim N(0,
1),\,\,\,\,\,\,\,\,
\frac{\sum_{i=m_1-ac+1}^{m_1}(1+A)Z_{i}}{cg(0)}\to 0
\end{eqnarray*}
as $c\to \infty$. Thus
\begin{eqnarray*}
E_v(T_c(g))\leq
m_1[1+\sum_{k=1}^{\infty}[\Phi(1+o(1))]^k]=\frac{(cg(0))^2}{\sigma^2}\frac{(1+o(1))}{1-\Phi(1)},
\end{eqnarray*}
where $\Phi(.)$ is the standard normal distribution.

Let $m_2=(cg(0))^2/(8\sigma^2 \ln c)$. Note that
\begin{eqnarray*}
&&\prod_{n=1}^{ac}\prod_{k=1}^nP_v\Big(\sum_{i=n-k+1}^{n}Z_{i}+c|g'(0)|n^{-1}\sum_{i=1}^{n}Z_{i}<cg(0)(1+o(1))\Big)\\
&&\geq
[P_v\Big(\sum_{i=1}^{ac}Z_{i}+\sum_{i=1}^{ac}Z'_{i}<cg(0)(1+o(1))\Big)]^{(ac)^2}=(1+o(1))(1-\Phi(\frac{\sqrt{c}g(0)}{\sqrt{a}(1+A)}))^{(ac)^2}
\to 1
\end{eqnarray*}
as $c\to \infty$, since
\begin{eqnarray*}
P_v(\frac{\sum_{i=1}^{ac}Z_{i}+\sum_{i=1}^{ac}Z'_{i}}{\sigma
(1+A)\sqrt{ac}}) \Rightarrow X \sim N(0, 1)
\end{eqnarray*}
as $c\to \infty$.  It follows  that
\begin{eqnarray*}
&&E_v(T_c(g))\geq \sum_{n=0}^{m_2}P_v(T_c(g)> n)\geq   m_2P_v(T_c(g)> m_2)\\
&\geq & m_2(1+o(1))\prod_{n=ac+1}^{m_1}\prod_{k=1}^nP_v\Big(\sum_{i=n-k+1}^{n}Z_{i}+\sum_{i=n-ac+1}^{n}Z'_{i}<cg(0)(1+o(1))\Big)\\
&\geq &  m_2(1+o(1))[P_v\Big(\sum_{i=1}^{m_2}Z_{i}+\sum_{i=m_2-ac+1}^{m_2}Z'_{i}<cg(0)(1+o(1))\Big)]^{m^2_2}\\
&=&m_2(1+o(1))[P_v\Big(\frac{\sum_{i=1}^{m_2}Z_{i}+\sum_{i=m_2-ac+1}^{m_2}Z'_{i}}{\sqrt{m_2}\sigma}<\sqrt{8\ln c}(1+o(1))\Big)]^{m^2_2}\\
&=&m_2(1+o(1))[\Phi(\sqrt{8 \ln
c})]^{m^2_2}=m_2(1+o(1))[1-\frac{1}{c^4\sqrt{8 \ln c}}]^{m^2_2}\to
m_2(1+o(1))
\end{eqnarray*}
as $c\to \infty$, where the third inequality comes from Theorem 5.1
in Esary, Proschan and Walkup (1967). Thus, we have
\begin{eqnarray*}
\frac{(cg(0))^2}{8\sigma^2 \ln c}(1+o(1))\leq E_v(T_c(g))\leq
\frac{(cg(0))^2}{\sigma^2(1-\Phi(1))}(1+o(1)).
\end{eqnarray*}

Let $v\in V^+$ and let
\begin{eqnarray*}
T_0=\min\{n: \sum_{i=1}^{n}Z_i+\frac{ac}{(ac)\wedge
n}\sum_{i=n-(ac)\wedge n+1}^{n}Z'_{i} \geq c\}.
\end{eqnarray*}
The uniform integrability of $\{T_c(g)/c\}$ for $c\geq 1$, follows
from the well-known uniform integrability of $\{T_0/c\}$ (see Gut
(1988)).

By the Strong Large Number Theorem we have
\begin{eqnarray*}
\mu &=&\lim_{n\to \infty}\frac{\sum_{i=1}^{n}Z_{i}}{n}\\
&=& \lim_{n\to \infty}\max_{1\leq j\leq
n}\frac{1}{n}[\sum_{i=j}^nZ_i+
g'(\mu)a^{-1}\sum_{i=n-ac+1}^{n}(Z_{i}-\mu)]
\end{eqnarray*}
Note that $T_c(g) \to \infty$ as $c\to \infty$,
\begin{eqnarray*}
\max_{1\leq j\leq T_c(g)}[\sum_{i=j}^{T_c(g)}Z_i
+g'(\mu)a^{-1}\sum_{i=T_c(g)-ac+1}^{T_c(g)}(Z_{i}-\mu)]\geq
cg(\mu)(1+o(1)),
\end{eqnarray*}
and
\begin{eqnarray*}
\max_{1\leq j\leq T_c(g)-1}[\sum_{i=j}^{T_c(g)-1}Z_i
+g'(\mu)a^{-1}\sum_{i=T_c(g)-ac}^{T_c(g)-1}(Z_{i}-\mu)]\leq
cg(\mu)(1+o(1)).
\end{eqnarray*}
It follows  that
\begin{eqnarray*}
&&\mu \longleftarrow \max_{1\leq j\leq T_c(g)}\frac{1}{T_c(g)}[\sum_{i=j}^{T_c(g)}Z_i +g'(\mu)a^{-1}\sum_{i=T_c(g)-ac+1}^{T_c(g)}(Z_{i}-\mu)]\\
&\geq & \frac{cg(\mu)(1+o(1))}{ T_c(g)}\\
& \geq  &\max_{1\leq j\leq
T_k-1}\frac{1}{T_c(g)-1}[\sum_{i=j}^{T_c(g)-1}Z_i
+g'(\mu)a^{-1}\sum_{i=T_c(g)-ac}^{T_c(g)-1}(Z_{i}-\mu)]
\longrightarrow \mu
\end{eqnarray*}
as $c\to \infty$. By the  uniform integrability of $\{T_c(g)/c\}$
and using Theorem A.1.1 in  Gut's book (1988), we have
\begin{eqnarray*}
E_v(T_c(g))=(1+o(1))\frac{cg(\mu)}{\mu}
\end{eqnarray*}
for a large $c$. This completes the proof of Theorem 2.

\bigskip

\textbf{Proof of Theorem 4.} Since $g(x)<0$ for $x>a^*$, $a^*\leq \mu^*$ and $\mu^*\geq 0$, it follows that
\begin{eqnarray*}
P_v\Big( m\hat{Z}_m <  cg(\hat{Z}_m),\,\,\,\hat{Z}_m > a^* \Big)\leq
P_v( \hat{Z}_m<\mu^*)
\end{eqnarray*}
and
\begin{eqnarray*}
P_v(T_c(g)>m)&=&P_v\Big(\sum_{i=n-k+1}^{n}Z_i <  cg(\hat{Z}_n), \,\,\, 1\leq k\leq n, \,\,1\leq n\leq m\Big)\leq P_v\Big(m\hat{Z}_m <  cg(\hat{Z}_m)\Big)\\
&= & P_v\Big(m\hat{Z}_m <  cg(\hat{Z}_m),\,\,\hat{Z}_m \leq  a^* \Big) +P_v\Big( m\hat{Z}_m <  cg(\hat{Z}_m),\,\,\,\hat{Z}_m > a^* \Big)\\
&\leq & 2P_v( \hat{Z}_m<\mu^*).
\end{eqnarray*}
Furthermore,
\begin{eqnarray*}
P_v( \hat{Z}_m<\mu^*)&=&P_v(\sum_{i}^{m}-Z_i >-m\mu^*)=P_v(\sum_{i}^{m}(\mu-Z_i) >m(\mu-\mu^*))\\
 &=& P_v(e^{\theta\sum_{i}^{m}(\mu-Z_i)} >e^{\theta m(\mu-\mu^*)})\leq e^{-m[\theta (\mu-\mu^*)-\ln M(\theta)]},
\end{eqnarray*}
where $M(\theta)=E_v(e^{\theta(\mu-Z_1)})$ and the last inequality
follows from Chebychev's inequality. Note that $h(\theta)=\theta
(\mu-\mu^*)-\ln M(\theta)$ attains its maximum  value
$h(\theta^*)=\theta^*(\mu-\mu^*)-\ln M(\theta^*)>0$ at
$\theta=\theta^*>0$, where $h'(\theta^*)=0$. So,
\begin{eqnarray*}
E_v(T_c(g))=1+\sum_{m=1}^{\infty}P_v(T_c(g)>m)\leq
1+2\sum_{m=1}^{\infty}e^{-m[\theta^* (\mu-\mu^*)-\ln
M(\theta^*)]}=\frac{e^{\theta^* (\mu-\mu^*)-\ln
M(\theta^*)}+1}{e^{\theta^* (\mu-\mu^*)-\ln M(\theta^*)}-1}.
\end{eqnarray*}
Let $k>1$. It follows that
\begin{eqnarray*}
E_{vk}(T_c(g)-k+1)^+&=&\sum_{m=1}^{\infty}P_{vk}(T_c(g)>m+k-1, T_c(g)>k-1 )\\
&\leq & (a_0+1)(k-1)P_0(T_c(g)>
k-1)+\sum_{m\geq(a_0+1)(k-1)}^{\infty}P_{vk}(T_c(g)>m+k-1).
\end{eqnarray*}
Similarly, we have
\begin{eqnarray*}
&&P_{vk}(T_c(g)>m+k-1)\\
&=&P_{vk}\Big(\sum_{i=n-k+1}^{n}Z_i <  cg(\hat{Z}_n), \,\,\, 1\leq k\leq n, \,\,1\leq n\leq m+k-1\Big) \leq  2P_{vk}( \hat{Z}_{m+k-1}<\mu^*)\\
&=&2P_{vk}\Big( \sum_{i=k-1}^{m+k-1}(\mu-Z_i)+ \sum_{i=1}^{k-1}(\mu_0-Z_i)>m(\mu-\mu^*)+(k-1)(\mu_0-\mu^*)\Big)\\
&\leq & 2\exp\{-m\Big(\theta^*(\mu-\mu^*)-\ln
M(\theta^*)+\frac{k-1}{m}[\mu_0-\mu^*-\ln M_0(\theta^*)]\Big)\}\leq
e^{-mb}
\end{eqnarray*}
for $m\geq (a_0+1)(k-1)$, since
\begin{eqnarray*}
\theta^*(\mu-\mu^*)-\ln M(\theta^*)+\frac{k-1}{m}[\mu_0-\mu^*-\ln
M_0(\theta^*)]\geq b
\end{eqnarray*}
for $m\geq (a_0+1)(k-1)$. Thus,
\begin{eqnarray*}
E_{vk}(T_c(g)-k+1)^+&\leq &(a_0+1)(k-1)P_0(T_c(g)\geq k)+2\sum_{m\geq (a_0+1)(k-1)}^{\infty} e^{-mb}\\
&\leq & (a_0+1)(k-1)P_0(T_c(g)>\geq
k)+\frac{2e^{-(a_0+1)(k-1)b}}{1-e^{-b}}.
\end{eqnarray*}

\end{document}